\documentclass[11pt]{article}
\usepackage{moriond,epsfig}

\def\Journal#1#2#3#4{{#1} {\bf #2}, #3 (#4)}

\def\MN{{\em MNRAS} }
\def\AJ{{\em AJ} }
\def\APJ{{\em ApJ} }

\def\be{\begin{equation}}
\def\ee{\end{equation}}
\def\bea{\begin{eqnarray}}
\def\eea{\end{eqnarray}}

\begin{document}
\vspace*{4cm}
\title{THE 9C SURVEY: A DEEPER SOURCE COUNT AT 15 GHZ}

\author{ E.M.WALDRAM AND G.G.POOLEY }

\address{Astrophysics Group, Cavendish Laboratory, Madingley Road,\\
Cambridge CB3 0HE, UK}

\maketitle\abstracts{
Foreground radio sources are a major contaminant for centimetre-wave cosmic microwave background (CMB) measurements and the 9C survey was set up as part of the observing  strategy of the CMB telescope, the Very Small Array. Prior to this survey with the Ryle Telescope at 15 GHz there was no comparable high-frequency radio survey of any extent. Our first published source count reached a limit of $\approx25$ mJy but we have now surveyed some areas more deeply, to a completeness limit of $\approx5$ mJy. We present the results from this deeper survey.}

\section{The Survey}
The 9C survey with the Ryle Telescope (RT) at 15.2~GHz (Waldram {\it et al}.~\cite{wa}) has been designed specifically for the source subtraction technique of the CMB telescope, the Very Small Array (VSA) at 34~GHz (Taylor {\it et al}.~\cite{ta1}). As the first high-frequency radio survey to cover an appreciable area, it is, however, of much wider interest. It is important in studies of the radio source population and source evolution, as well as in predicting the effect of foreground sources on CMB observations over a range of wavelengths.

The prime purpose of the survey has been to define a catalogue of sources which must be monitored by the VSA during its observations and so the choice of fields and their depth in different areas are determined by the VSA observing programme. The fields were selected to be at least $25^\circ$ from the Galactic plane, to be spaced in RA, and to contain, as far as could be predicted from lower frequency surveys, no very bright ($> 0.5 $ Jy) radio sources at 34~GHz. (See Figure 1.) The depths of the survey areas depend on the sensitivity requirements of the VSA; we allow for the possibility of extremely inverted source spectra between 15 and 34~GHz, i.e. as far as $\alpha_{15}^{34} = -1$ (taking $S\propto \nu^{-\alpha}$). 

For the VSA compact array observations ($l = $ 150 to 800) we have covered an area of $\approx$~520~deg$^{2}$ to $\approx25$ mJy completeness (Taylor {\it et al}.~\cite{ta2}) and for the extended array ($l = $~300~to~1500) an area of $\approx190$~deg$^{2}$ to $\approx10$ mJy completeness (Grainge {\it et al}.~\cite{gr}). Since the the FWHM of the RT primary beam is only 6 arcmin, compared with the VSA's $4.6^\circ$ in its compact array and $2.0^\circ$ in its extended array, it has been necessary to develop a rastering technique for our RT observations in order to keep pace with the VSA schedule. (See Waldram {\it et al}.~\cite{wa}.) 

\section{The deeper areas}
The deeper survey described here has been set up for use with the on-going deeper VSA extended array observations and currently covers $\approx12$ deg$^{2}$ to $\approx5$ mJy completeness. To achieve this depth we observe approximately  0.3~deg$^{2}$ of sky in 12~h, with a raster scan of 72 different pointings, and integrate a series of 4 to 6 sets of such data. Maps from the individual pointings are CLEANed before combining them into the final `raster' map. In this way we reach a noise level of $< 1$ mJy. Peaks on the raster maps are followed up with pointed observations to establish the genuine sources with reliable flux densities. So far, we have a catalogue of 97 sources with 73 above the completeness limit.
The three areas surveyed in this way are in the regions RA~$00^{\rm h}20^{\rm m}$  Dec $+30^\circ$, RA $09^{\rm h}35^{\rm m}$ Dec $+31^\circ$ and RA $15^{\rm h}38^{\rm m}$ Dec $+45^\circ$ (B1950.0).
\begin{figure}
        \centerline{\epsfig{file=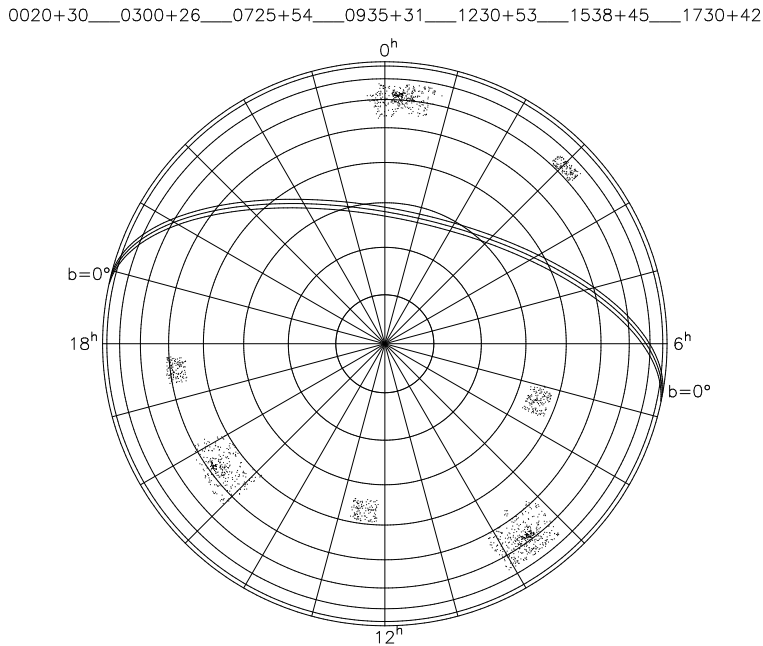,
        angle=270,
        width=10.0cm,clip=}} 
        \caption{The 9C survey regions: an equatorial plane projection, with the N. pole at the centre. Sources are shown as dots. The Dec. circles are at intervals of $10^\circ$ and the Galactic plane is shown. (B1950.0 coordinates).}
\end{figure}

\section{The Source Count}
In Figure 2 we show the original differential source count (Waldram {\it et al}.~\cite{wa}) for sources $> 25$~mJy, together with the function fitted to that count, which has the form:

\[
n(S) \equiv \frac{{\rm d}N}{{\rm d}S} \approx 51 \left( \frac{S}{\rm Jy} \right)^{-2.15}
\, {\rm Jy}^{-1}{\rm sr}^{-1}
\]

We have now added the source count from the deeper area and it can be seen that this lies on the same line, within the error bars. There is no indication of a turn-over in the count at this level. The predicted differential count from the model of Toffolatti {\it et al}.~\cite{to} is also shown.
 
\begin{figure}
        \centerline{\epsfig{file=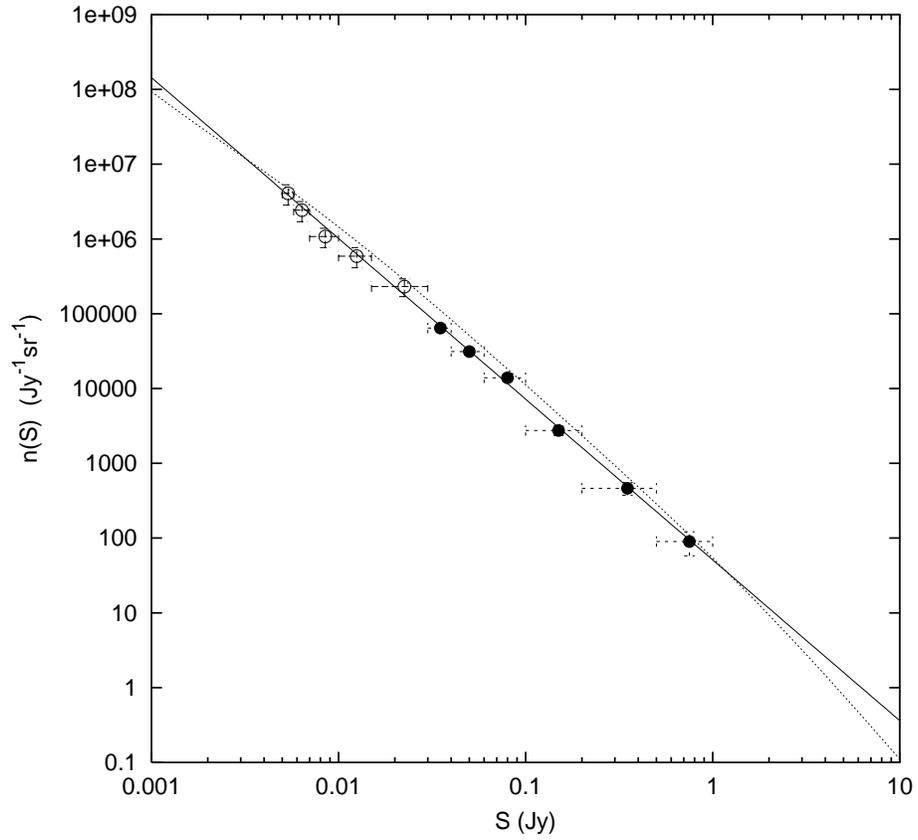,
        angle=270,
        width=12.0cm,clip=}} 
        \caption{Plot of the differential source count for both the original survey (filled circles) and the current deeper survey (open circles). The solid line is the function we fitted to the original count and the dotted line is the prediction from the model of Toffolatti et al. }
\end{figure}

\begin{figure}
        \centerline{\epsfig{file=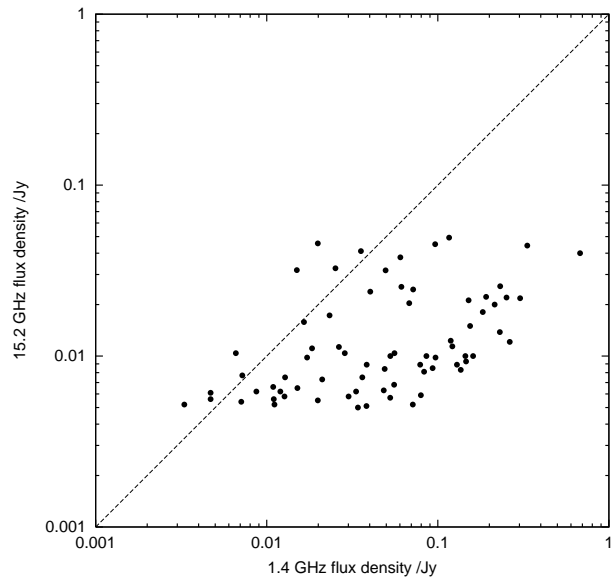,
        angle=270,
        width=8.0cm,clip=}} 
        \caption{Plot of 15.2~GHz flux density v. 1.4~GHz flux density for sources $\geq5$~mJy. The line indicates $\alpha_{1.4}^{15.2} = 0$ . }
\end{figure}

\begin{figure}
        {\epsfig{file=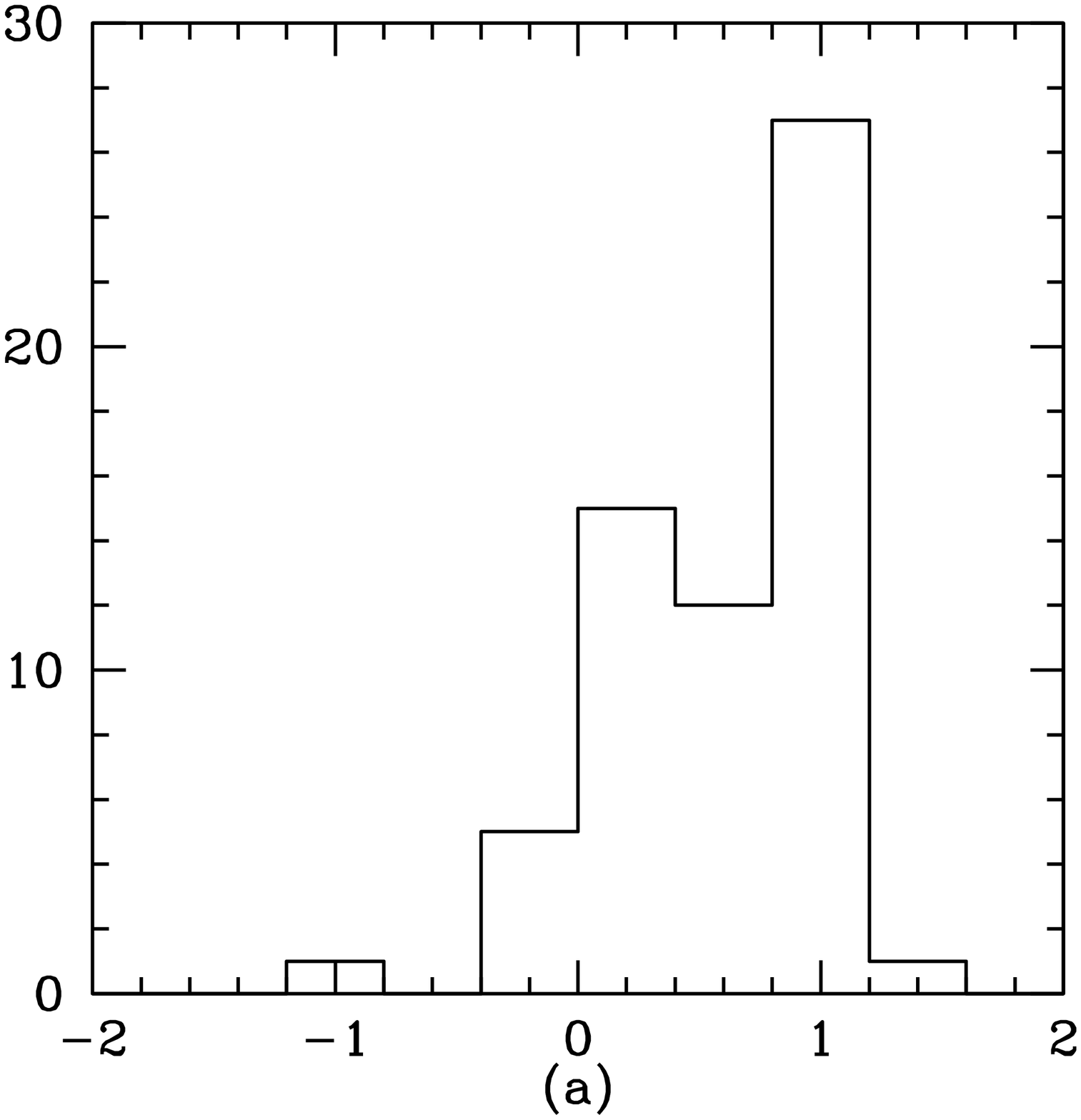,
        width=5.25cm,clip=}} 
        {\epsfig{file=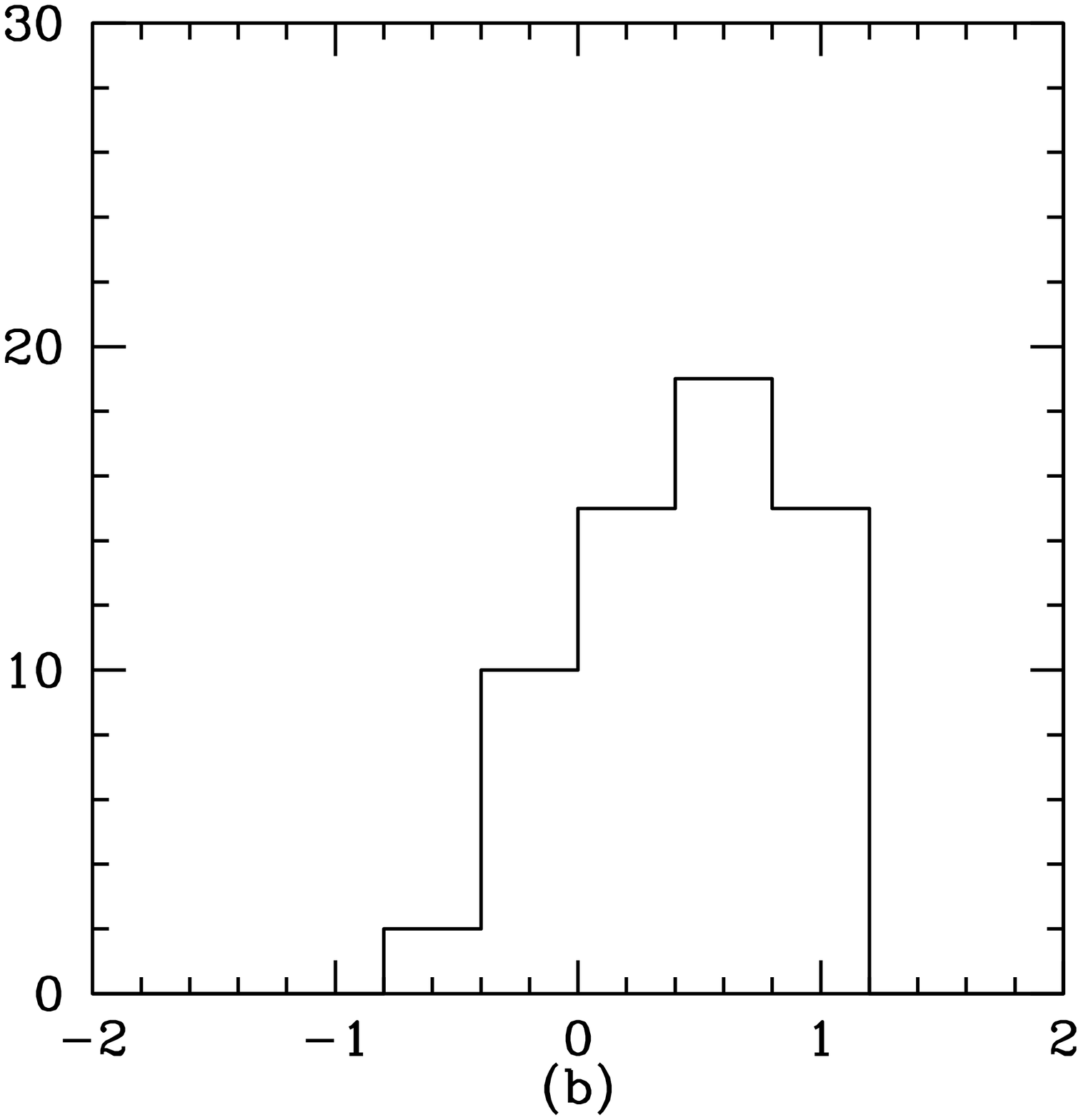,
        width=5.25cm,clip=}} 
        {\epsfig{file=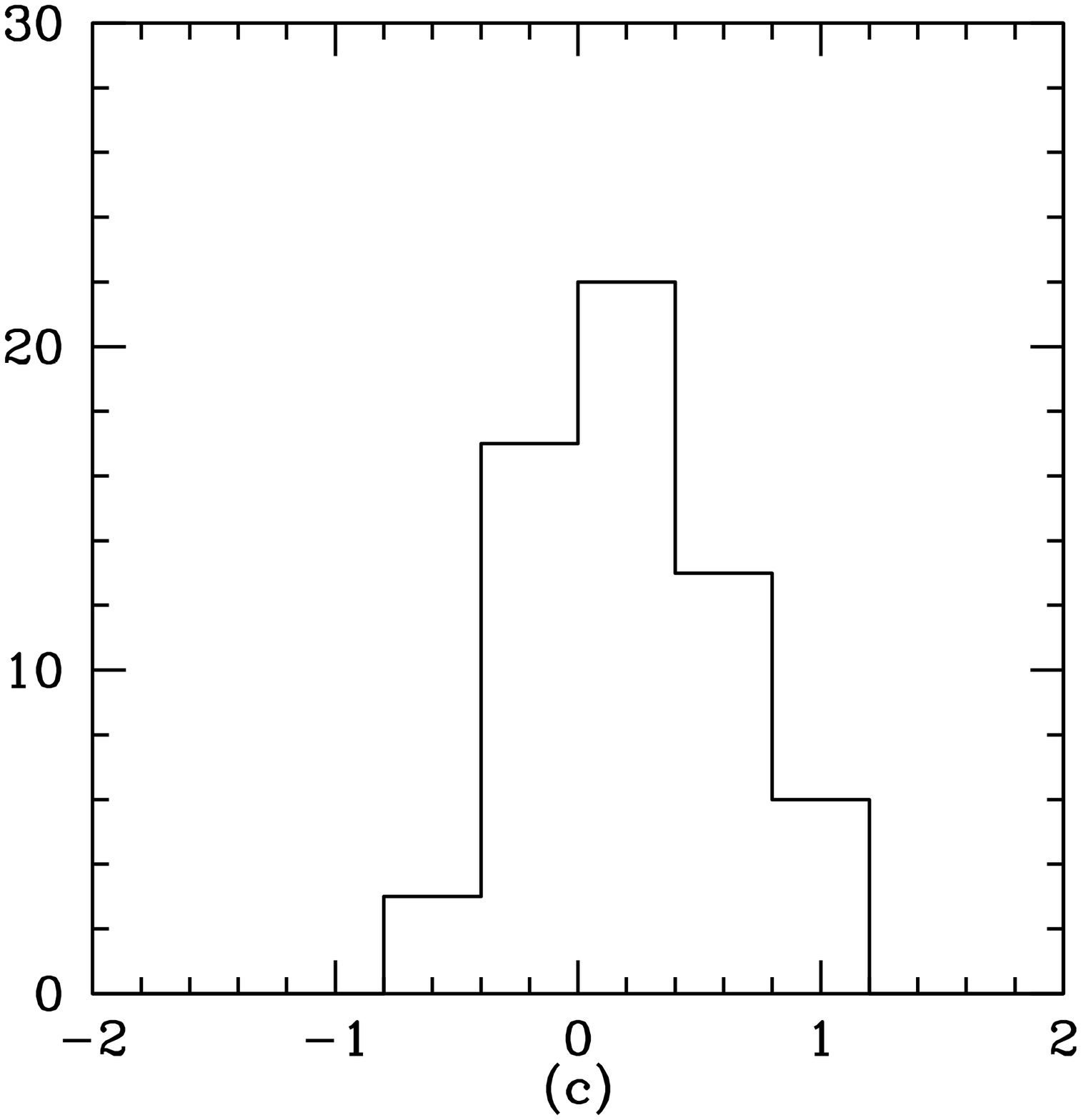,
        width=5.25cm,clip=}} 
        \caption{Distributions of spectral index $\alpha_{1.4}^{15.2}$ for complete samples each of 61 sources in the 15 GHz flux density ranges: (a) 5 to 25~mJy, (b) 25 to 100~mJy, (c) above 100~mJy (taking $S\propto \nu^{-\alpha}$). 
For the source shown at $\alpha = -1$ in (a) the value is an upper limit.}
\end{figure}

\section{Spectral Indices}
We have correlated our list of 73  sources $\geq5$~mJy with surveys at 1.4~GHz (see Figure 3) and find that only one source, 5.3 mJy at 15~GHz, is without a counterpart in either NVSS (Condon {\it et al}.~\cite{co}) or FIRST (Becker {\it et al}.~\cite{be}). In Figure 4 we show the spectral index distribution $\alpha_{1.4}^{15.2}$ for the 61 sources in the range 5 to 25 mJy and for comparison we have included plots for complete samples of the same number of sources in higher flux density ranges. It can be seen that there are apparent differences in the distributions, in particular in the numbers of sources with inverted spectra. In the range 5 to 25~mJy 10 per cent have $\alpha_{1.4}^{15.2} < 0$, in the range 25 to 100~mJy it is 20 per cent, and in the sample above 100~mJy 33 per cent. Further work is needed to investigate the significance of these results.

\section*{Acknowledgments}
We are grateful to the staff of our observatory for the operation of the Ryle Telescope, which is funded by PPARC, and to members of the VSA collaboration for useful discussions. We also thank Luigi Toffolatti for providing us with his predicted count in numerical form.

\end{document}